\documentclass{nature}

\usepackage{blindtext, graphicx,caption}
\usepackage{subfig}
\usepackage{wrapfig}
\usepackage{fontawesome}
\usepackage{array,paralist,hyperref,amssymb,amsmath,empheq,mathtools}
\usepackage{multirow,framed}

\usepackage[usenames, dvipsnames]{color}

\usepackage{pifont}
\usepackage{verbatim}
\usepackage{xcolor}
\usepackage{pgfplots}
\usepackage{pgfplotstable}
\pgfplotsset{compat=newest}
\usepackage{tikz}
\usetikzlibrary{shapes,arrows}
\usepackage{amsmath,bm,times}
\usepackage{csquotes}

\usetikzlibrary{chains,shapes.multipart}
\usetikzlibrary{shapes,calc}
\usetikzlibrary{automata,positioning,fit}
\usepackage{ifthen}
\ifthenelse{\isodd{\value{page}}}{}{}

\newcommand{\cmark}{\ding{51}}%
\newcommand{\xmark}{\ding{55}}%

\definecolor{ashgrey}{rgb}{0.7, 0.75, 0.71}
\definecolor{asparagus}{rgb}{0.53, 0.66, 0.42}
\definecolor{burlywood}{rgb}{0.87, 0.72, 0.53}
\definecolor{bondiblue}{rgb}{0.0, 0.58, 0.71}
\definecolor{babyblueeyes}{rgb}{0.68, 0.87, 0.98}

\usepackage{hyperref}
\hypersetup{
  colorlinks=false,
  pdfborder = {0 0 0}
}

\begin{document}
%
\title{SAFETY: Secure gwAs in Federated Environment Through a hYbrid solution with Intel SGX and Homomorphic Encryption}

\author{Md Nazmus Sadat$^{1}$, Md Momin Al Aziz$^{1}$, Noman Mohammed$^{1}$, Feng Chen$^{2}$, Shuang Wang$^{2}$, Xiaoqian Jiang$^{2}$}

%


%


\maketitle

\begin{affiliations}
 \item Department of Computer Science, University of Manitoba, Winnipeg, MB R3T 2N2, Canada
 \item Department of Biomedical Informatics, University of California San Diego, La Jolla, CA, 92093, USA
\end{affiliations}

\begin{abstract}
Recent studies demonstrate that effective healthcare can benefit from using the human genomic information. For instance, analysis of tumor genomes has revealed 140 genes whose mutations contribute to cancer~\cite{vogelstein2013cancer}. As a result, many institutions are using statistical analysis of genomic data, which are mostly based on genome-wide association studies (GWAS). GWAS analyze genome sequence variations in order to identify genetic risk factors for diseases. These studies often require pooling data from different sources together in order to unravel statistical patterns or relationships between genetic variants and diseases. In this case, the primary challenge is to fulfill one major objective: accessing multiple genomic data repositories for collaborative research in a privacy-preserving manner. Due to the sensitivity and privacy concerns regarding the genomic data, multi-jurisdictional laws and policies of cross-border genomic data sharing are enforced among different regions of the world. In this article, we present SAFETY, a hybrid framework, which can securely perform GWAS on federated genomic datasets using homomorphic encryption and recently introduced secure hardware component of Intel Software Guard Extensions (Intel SGX)~\cite {hoekstra2013using} to ensure high efficiency and privacy at the same time. Different experimental settings show the efficacy and applicability of such hybrid framework in secure conduction of GWAS. To the best of our knowledge, this hybrid use of homomorphic encryption along with Intel SGX is not proposed or experimented to this date. Our proposed framework, SAFETY is up to 4.82 times faster than the best existing secure computation technique.
\end{abstract}



\section{Introduction}\label{intro}
Rapid advancement in human genome sequencing has led us to a genomic era where human genomic data play an ever-important role in clinical research~\cite{burke2007personalized}. As cost-effective and efficient genome sequencing technologies are readily available, the research community can conduct experiments on different genomic data repositories for scientific discovery~\cite{brenner2013prepared}. As a result of this massive data availability, Genome-Wide Association Studies (GWAS) are gaining popularity as they answer critical questions like susceptibility towards a disease or a physical trait by analyzing genome sequence variations in different individuals. GWAS examine genetic architecture of a disease to identify genetic risk factors associated with it. In other words, GWAS aims at finding if there are any correlations between a certain gene and a specific disease. Another fundamental goal of GWAS is to identify biological factors responsible for disease susceptibility in order to develop more effective diagnosis, treatment, and prevention techniques.
\par
A larger genomic dataset is quintessential to perform any analytical study such as GWAS. Different research organizations or healthcare facilities often sequence genomes of different patients or participants for this reason. Researchers are interested in executing queries over these massive genomic datasets for unraveling new pieces of information about diseases under study. Oftentimes, the accuracy of this evaluation relies on the quantity and quality of the data used in the analysis-- but a single organization often does not possess adequate genomic data (collection, processing, and storing of large-scale data is non-trivial) to perform a comprehensive or meaningful experiment.  Because more data can reduce the sampling errors and improve the power of the analysis (for instance, statistical strength of GWAS increases with the quantity of data~\cite{visscher2012five}), organizations tend to collect as much data as possible to meet data analysis needs.
\par
Because sharing genomic data in plaintext possesses serious privacy implications for the participants \cite{erlich2014routes}, in addition to the approval from an institutional review board (IRB), collaborative research on shared genomic data often needs to satisfy two criteria at the same time --- a) authorizing access to  genomic data for research and b) preserving participants' privacy and protecting the confidentiality of their genomic information ~\cite{canadahealthdata}. That is why strict policies regarding genomic data sharing have been enforced, and generally, these policies are different in different regions of the world. This difference in the regulations of cross-border genomic data sharing greatly impedes international research projects \cite{hayden2013geneticists}. It is imperative to address the reality challenge with practical solutions to promote health science discoveries.   

\subsection{Contributions.} In this paper, we propose a hybrid framework, SAFETY, for secure execution of some popular statistical tests used in GWAS in a federated environment. Our proposed hybrid model incorporates security and efficiency of two different cryptographic schemes in a single system. More precisely, it is the first attempt to infuse homomorphic encryption with SGX to develop a secure and scalable genomic data computation model. The experimental results clearly demonstrate that it performs consistently \textit{irrespective of the number of data owners making it highly scalable} (see Section \ref{results} for details). This hybrid model captures the essence of both techniques: ability of computing some functions on encrypted data (homomorphic encryption) and performing sophisticated mathematical operation in the secure execution area of a SGX enabled CPU. Our primary goal behind proposing a hybrid model relies on better security guarantee from existing secure computation schemes, and faster and scalable execution for any number of data owners. From our experimental results, it is evident that the proposed hybrid model provides better efficiency and security than pure secure hardware or homomorphic encryption based solutions. 

\par
\begin{figure}
\centering
\captionsetup{justification=centering,margin=.5cm}
    \includegraphics[width=\textwidth]{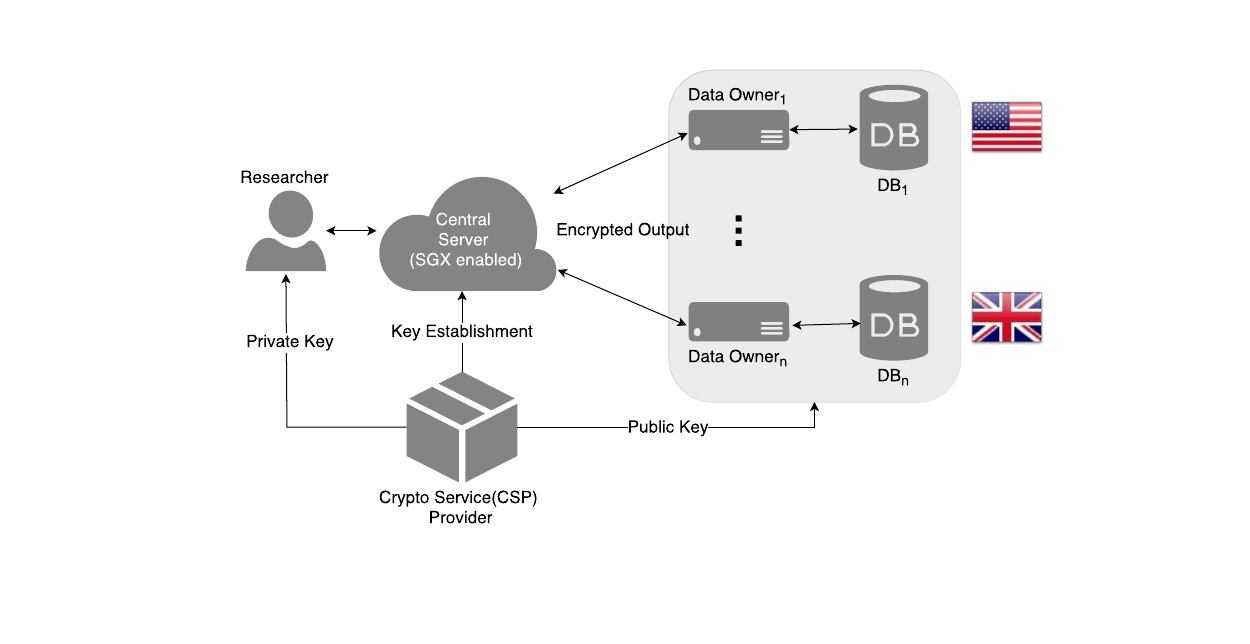}
    \caption{Block diagram of the federated architecture where data owners are geographically distributed.} 
    \label{fig:architechture}
\end{figure}

 SAFETY utilizes an architecture~\cite{aziz2016secure} to execute secure count query on federated genomic datasets. Similar federated architectures are available in literature~\cite{constable2015privacy,bogdanov2014privacy}. 
In our adopted architecture~\cite{aziz2016secure,zhang2015foresee}, genomic data resides in the local premises of individual data owners in plaintext (see Figure \ref{fig:architechture}). Data owners have their own database systems which are geographically distributed and have different policy compliance for the data usage. An overview of data representation for each data owner is shown in Table \ref{table:data_represent}. Proper authentication allows any researcher to execute queries on their data. 
\par 
Among the existing secure computation techniques, SGX is most efficient (existing techniques are described in supplementary document). For instance, an implementation of SGX-based MapReduce framework \cite{schuster2015vc3} shows a very modest overhead of $8\%$ to achieve read/write integrity. This is a great advantage of SGX in comparison to other secure computation schemes like garbled circuit and homomorphic encryption, which generally increase the computational overhead thousands of times \cite{chen2017princess}. However, our proposed hybrid model is \textit{1.7} to \textit{4.82} times faster than SGX (see Section \ref{results}). This comparative efficiency increases with the number of data owners. The contributions of this article are summarized as follows:
    \begin{enumerate}
        \item  We propose a hybrid cryptographic framework, SAFETY, which uses homomorphic encryption along with secure hardware features of the Intel SGX. SAFETY is not only secure and efficient, but also overcomes the limitations of solely homomorphic encryption based solutions which often come with higher computational overhead for processing higher order polynomials. In addition, SAFETY also simplifies solely SGX based solutions, which require pairwise attestation and secure key distributions between server and data owners.
        \item Using SAFETY, we securely execute and evaluate some of the major functions of GWAS in federated architecture where genomic data are distributed and owned by different parties. We performed four statistical tests: Linkage Disequilibrium (LD), Hardy-Weinberg Equilibrium (HWE), Cochran-Armitage Test for Trend (CATT), Fisher's Exact Test (FET) to evaluate SAFETY over a variety of settings. However, our framework SAFETY can incorporate any GWAS functions (i.e., transmission disequilibrium test~\cite{chen2017princess}, EigenSTRAT~\cite{price2006principal}, linear mixed model~\cite{simmons2016enabling}, etc.) and not limited to the GWAS functions mentioned previously. The methodology to perform these statistical tests securely is discussed in Section \ref{sec:methods}.
        \item SAFETY ensures that each data owner is completely unaware of the contributions from the other data owners, who are participating in the same analysis. Moreover, the final result is revealed only to the researchers without disclosing individual contribution of data owners. This allows us to preserve the privacy of the output of each data owner.
        \item We conduct multiple experiments in different realistic setting in a federated environment varying the data size and the geographic locations of data owners (see Section \ref{results} for details).
    \end{enumerate}

\section{Results}
\label{results}
In this section, we extensively evaluate aforementioned hybrid approach and secure hardware based approach in a federated environment using Amazon cloud and demonstrate their applicability in a real world setting. Our proposed framework SAFETY is based on hybrid approach where we use SGX along with homomorphic encryption. However, secure hardware based approach uses only SGX. See Section \ref{sec:methods} for more details.
\subsection{System Architecture}
The proposed system as shown in Figure~\ref{fig:architechture} has  four main entities: 
\begin{itemize}

    \item \textit{Researchers (authorized):} Individuals or organizations who want to execute queries over genomic databases. This party sends their queries to the central server and expect different encrypted results of different GWAS functions.

    \item \textit{Data Owners:} These parties are geographically distributed and possess databases upon which queries are performed. Data owners might be hospitals or government organizations who want to share their genomic data and have different policies regarding the data sharing. The proposed model supports any number of data owners where they can execute any aggregate query locally.

    \item \textit{Crypto Service Provider (CSP):} It manages the cryptographic keys that will be used for data encryption and decryption in different stages of our system. Each data owner receives a public key from the \textit{CSP} and uses it to encrypt its outputs. CSP also issues the private key to an authorized researcher who can decrypt the final result.

\item \textit{Central Server:} The central server communicates with all the other entities. It receives queries from the researcher, forwards them to all data owners and collects individual encrypted results from each data owners. Individual encrypted results from data owners are securely combined by the central server to compute the final result of the query with the help of homomorphic encryption and SGX. 
\end{itemize}
\begin{table}
    \begin{center}
    \caption{Data representation in each party\label{table:data_represent}}
    \resizebox{.6\textwidth}{!}{\begin{tabular}{ |c|c|c|c|c|c|c| }
        \hline
        \, &\, & \multicolumn{4}{c|}{\textbf{Sequence}} & \,\\
        \hline
         \# & \textbf{Case} & \textbf{rs4426} & \textbf{rs4305} & \textbf{rs4630} & \, & \textbf{Cancer}\\
        \hline
        \multirow{3}{4em}{Data Owner 1} & 1 & CC &CT &GG & \ldots & Negative  \\
        & 2 & CT &CT &AG & \ldots & Negative  \\
        & 3 & CC &CT &GG & \ldots & Negative  \\
        \hline
        \multirow{3}{4em}{Data Owner 2} & 1 & CC &CT &GG & \ldots & Negative  \\
        & 2 & CT &CC &GG & \ldots & Positive  \\
        & 3 & CC &CT &GG & \ldots & Positive  \\
        \hline
        \multirow{3}{4em}{Data Owner 3} & 1 & CT &CC &AG & \ldots & Positive  \\
        & 2 & CT &CT &AG & \ldots & Negative  \\
        & 3 & TT &CC &GG & \ldots & Positive  \\
        \hline
        \multirow{3}{4em}{Data Owner 4} & 1 & TT &CC &AA & \ldots & Positive  \\
        & 2 & CC &CC &GG & \ldots & Positive  \\
        & 3 & CC &CT &GG & \ldots & Positive  \\
        \hline
    \end{tabular}}
    \end{center}
\end{table}

\subsection{Experimental Settings}
\begin{table}[ht]
    \centering
    \caption{Server locations and average latency\label{table:ping}}
    \begin{tabular}{ |c|c|c| } 
     \hline
     \textbf{Server Location} & \textbf{IP Address} & \textbf{Network Latency (ms)} \\
            \hline
            Canada (Manitoba) & 130.179.30.133 & $<$1\\
            USA (Oregon) & 52.32.83.223 & 37\\
            London & 52.56.65.221 & 105\\
            Seoul & 52.78.100.194 & 170\\
            Sydney & 54.206.67.251 & 233\\  
     \hline
    \end{tabular}
\end{table}
In our experimental setup, the researcher, CSP, and central server were located in Manitoba, Canada. Our central server was hosted on a machine with Intel Core i7-6700 (3.40 GHz) processor and 8 GB memory. However, we emulate data owners in different locations of the world to evaluate the propriety of our proposed framework in a real world environment. We used \textbf{\textit{Amazon EC2 cloud servers}} having the same configuration for all data owners. Table \ref{table:ping} shows the location, IP address, and the latency of these servers used in our experiment. In our experiments, we used 80 bit security (size of the public key is 1024 bits) on the public-key cryptosystem. The security can be improved by increasing the key length.
\par
We performed four experiments with different settings. The number of data owners was different for different experiments which allowed us to evaluate the scalability of both methods. Table \ref{table:experiments} shows different settings used in the experiments. For instance, in experiment 1, two data owners were in USA and Canada while in experiment 4, five data owners were residing in all the locations mentioned in Table \ref{table:ping}. Experiments were performed using synthetic data which were generated according to the allele frequency of CHB, CHS, JPT and MXL populations from \textit{1000genomes} dataset (August 2010 Release) \cite{1000genomephase1}. 
\begin{table}[t]
    \begin{center}
    \caption{Location of different data owners in different experimental settings\label{table:experiments}}
    {\begin{tabular}{ |p{1.8cm}|c|c|c|c|c|c| }
    \hline
    \textbf{Exp. \#}  & \textbf{Canada}  & \textbf{USA}  & \textbf{London}  & \textbf{Seoul} & \textbf{Sydney} \\[3ex]
    
    \hline
    Exp. 1  & \cmark  & \cmark& \xmark& \xmark & \xmark\\
    \hline
     Exp. 2 & \cmark  & \cmark& \cmark& \xmark & \xmark\\
    \hline
    Exp. 3  & \cmark  & \cmark& \cmark& \cmark & \xmark\\
    \hline
    Exp. 4  & \cmark  & \cmark& \cmark& \cmark & \cmark\\
    \hline
    \end{tabular}}
    \end{center}
\end{table}

\subsection{Experimental Results}\label{subsec:exp_result}
Prior to analyzing the running times of our proposed methods, we evaluate the required time to compute the four statistical tests on plaintext (i.e. without any security protection). We calculate the exact results for the GWAS calculations without loosing any accuracy.
\par
\begin{figure}
\centering
\captionsetup{justification=centering,margin=.5cm}
    \includegraphics[width=0.7\textwidth]{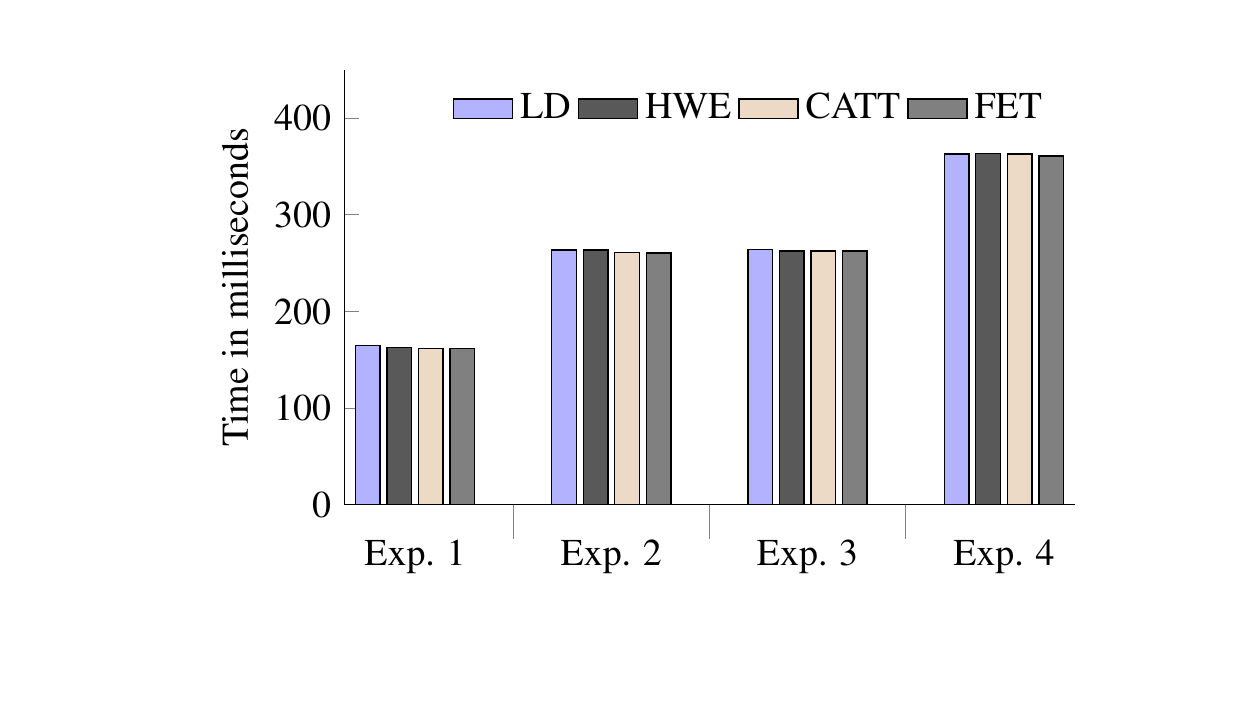}
    \caption{Experimental results for plaintext.} \label{fig:resultPlain}
\end{figure}

In Figure \ref{fig:resultPlain}, we show the running time (in milliseconds) for performing the four statistical tests on plaintext. We observed that in any single experimental setup, the running time is almost the same for all the statistical tests. However, running times for different experiments are different because different experiments involve different number of data owners (as shown in Table \ref{table:experiments}). As a result, higher communication overhead is added in these experiments. For instance, experiment 2 involves more data owners than experiment 1, which yields more communication overhead and results in greater running time. 
\par
\begin{figure}
\centering
\captionsetup{justification=centering}
    \includegraphics[width=\textwidth]{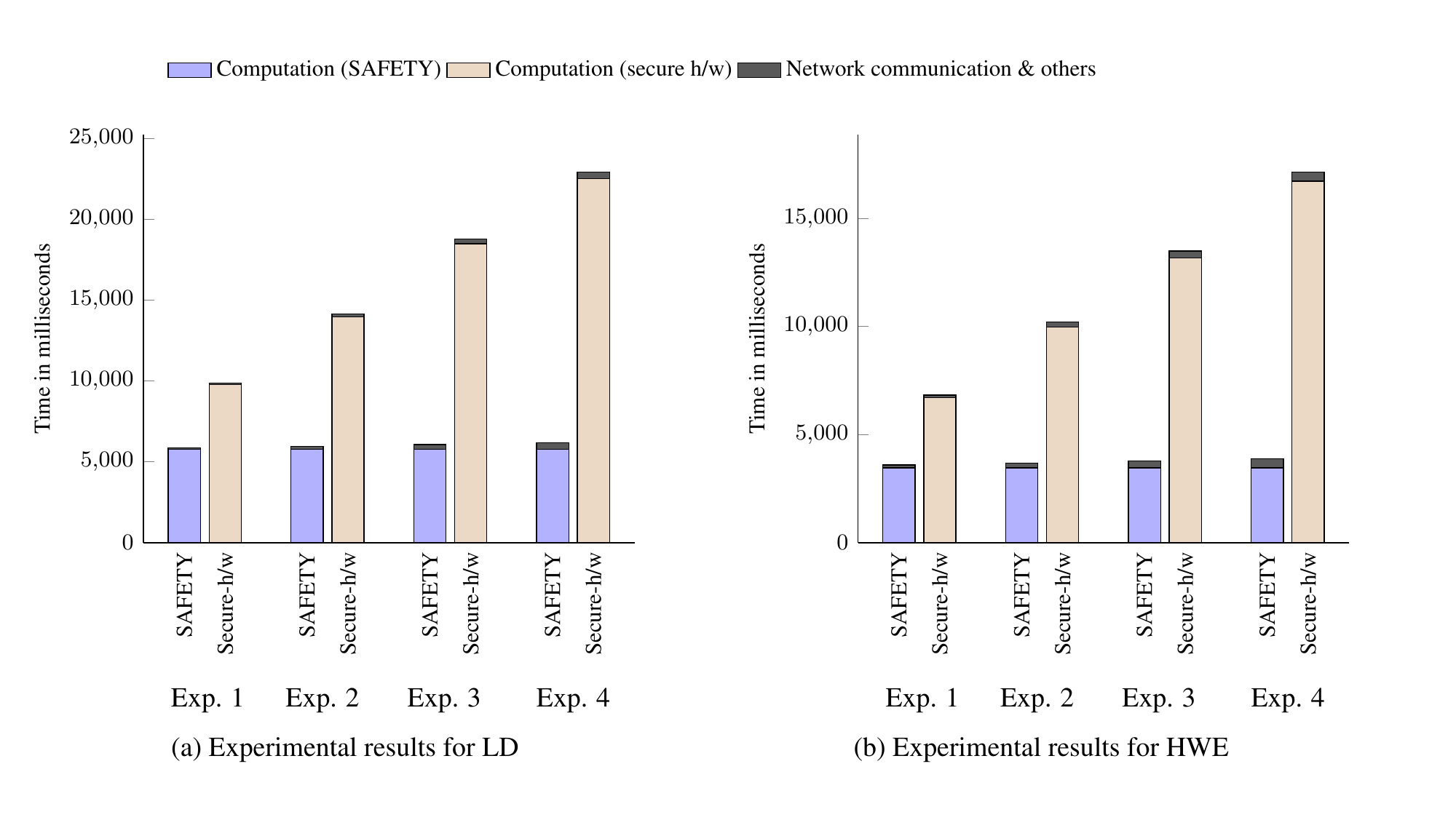}
    \caption{Experimental results for LD and HWE.  (a) and (b) compared computation time and communication costs of conducing Linkage Disequilibrium (LD) and Hardy-Weinberg Equilibrium (HWE) test using SAFETY and purely secure hardware approach.}
\label{fig:exp_ld_hwe_cluster}
\end{figure}
\begin{figure}[t]
\centering
\captionsetup{justification=centering,margin=.5cm}
    \includegraphics[width=\textwidth]{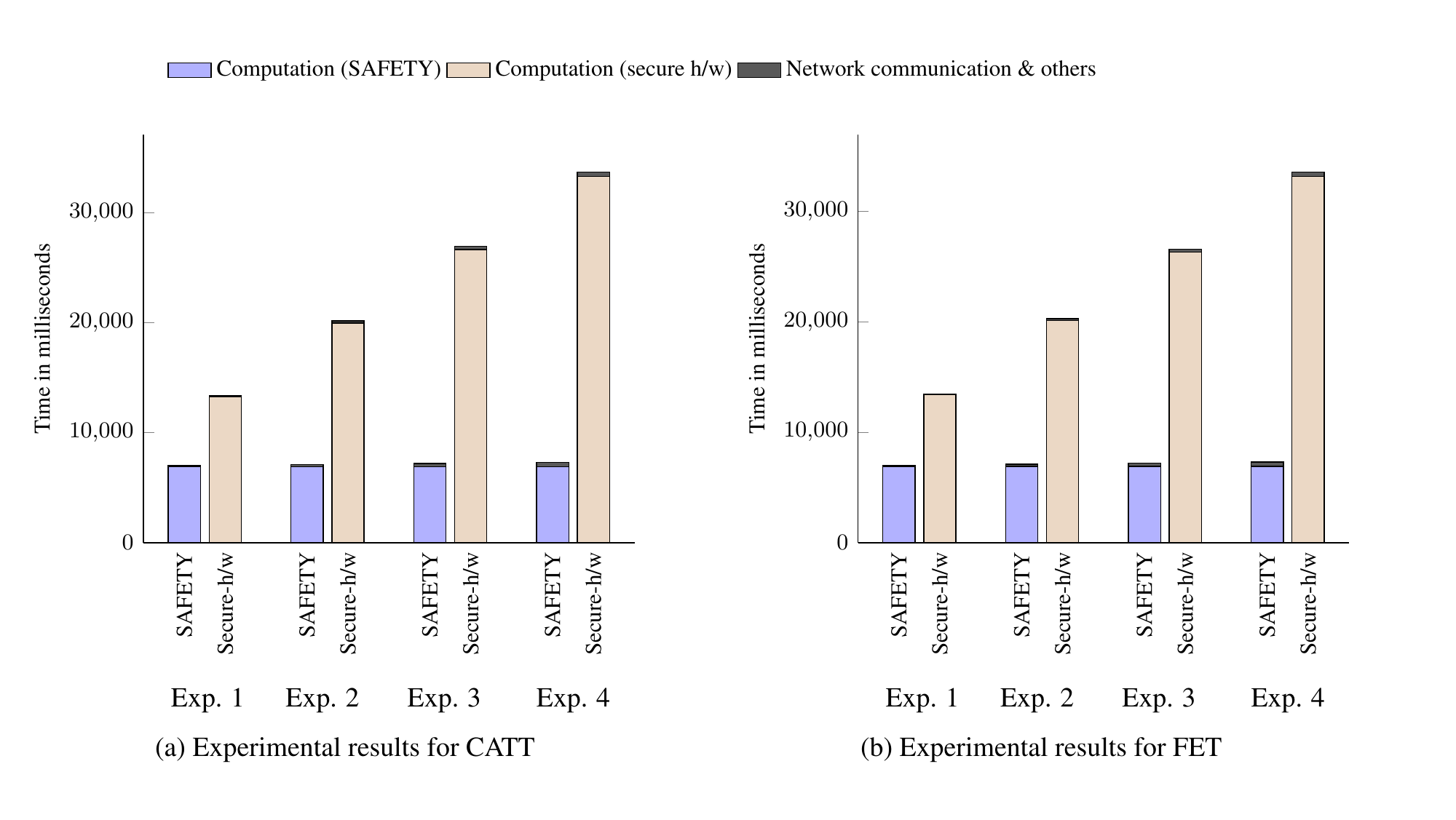}
    \caption{Experimental results for CATT and FET. (a) and (b) compared computation time and communication costs of conducing Cochran-Armitage Test for Trend (CATT) and Fisher’s Exact Test (FET) using SAFETY and purely secure hardware approach.}
\label{fig:exp_catt_fet_cluster}
\end{figure}

The running time for computing LD on ciphertexts is shown in Figure \ref{fig:exp_ld_hwe_cluster} (a). Here, running time is decomposed into communication overhead in the network and time required for secure computation of the method. It is noteworthy that,
\begin{equation*}
    \text{Communication overhead }  \propto \text{ Number of data owners }
\end{equation*}
SAFETY requires 5,770 ms to compute LD coefficient for two data owners, which is 1.7 times faster than secure hardware approach. Rest of the time is due to the communication overhead in the network. Figure \ref{fig:exp_ld_hwe_cluster}(b), \ref{fig:exp_catt_fet_cluster}(a), and \ref{fig:exp_catt_fet_cluster}(b) illustrate the experimental results for performing HWE, CATT, and FET respectively on ciphertexts. Experimental results illustrate that SAFETY is much faster than solely secure hardware based approach. For instance, for HWE, SAFETY is 1.93, 2.87, 3.8, and 4.82 times faster than solely secure hardware based approach in Experiment 1, 2, 3, and 4 respectively (see Figure \ref{fig:exp_ld_hwe_cluster}(b)).  It is noteworthy that SAFETY and the secure h/w approach both utizlize the asymmetric encryption and decryption.

\par 
The experimental results demonstrate that the performance of the secure hardware approach does not scale well with the number of data owners. As the number of data owner increases, the running time of secure hardware approach increases rapidly. On the contrary, the hybrid approach (SAFETY) performs consistently irrespective of the number of data owners due to its hybrid properties (homomorphic addition followed by computation inside enclave). In this case, only the communication overhead increases which is very small considering the total running time.

\par 
Another important analysis regarding the methods is the number of decryptions needed for any statistical tests. It is evident that LD requires more time than HWE while CATT and FET require more time than the other two. The reason behind this is, the time required to perform a statistical test is proportional to the number of decryptions required. Moreover, for secure hardware approach, all the individual contributions of data owners need to be decrypted inside the secure enclave. As the number of the data owner increases, the number of decryptions also increases which results into higher running time. Table \ref{table:decryptionCount} demonstrates how the required number of decryptions increases with the number  of data owners. 
\begin{table}[ht]
\centering
\caption{Number of decryptions required to perform a statistical test for different number of data owners \label{table:decryptionCount}}
\begin{tabular}{|c|c|c|c|c|c|c|}
\hline
\multirow{3}{*}{\textbf{Test}} & \multicolumn{6}{c|}{\textbf{Number of data owners}}                                        \\ \cline{2-7} 
                                  & \multicolumn{2}{c|}{One} & \multicolumn{2}{c|}{Three} & \multicolumn{2}{c|}{Five} \\ \cline{2-7} 
                                  & Hybrid    & Secure h/w   & Hybrid     & Secure h/w    & Hybrid    & Secure h/w    \\ \hline
LD                                & 4         & 4           & 4          & 12            & 4         & 20            \\ \hline
HWE                               & 3         & 3            & 3          & 9            & 3         & 15            \\ \hline
CATT                              & 6         & 6           & 6          & 18            & 6         & 30            \\ \hline
FET                               & 6         & 6           & 6          & 18            & 6         & 30            \\ \hline
\end{tabular}
\end{table}



\section{Methods}\label{sec:methods}
In this section, at first, we introduce some of the concepts, which are required to understand the proposed methods. Then, we present the threat model. Finally, we discuss how to perform the statistical tests (LD, HWE, CATT, and FET) securely. Please see the supplementary document for a brief introduction of the corresponding GWAS functions. To explain our proposed methods we use the data from Table \ref{table:data_represent}.\\
As mentioned earlier, we consider the use of Intel SGX in two ways ---
\begin{inparaenum} [\itshape 1\upshape)] 
\item Hybrid approach: using Intel SGX along with homomorphic encryption 
\item Secure hardware approach:  using only Intel SGX.
\end{inparaenum}
SAFETY is based on the hybrid approach.

\subsection{Intel SGX}\label{sgx}
Intel SGX is a set of extensions to the Intel architecture which mainly focuses on the problem of running applications on a remote machine administered by an untrusted party. SGX allows parts of an application to be executed inside secure segments of the CPU called \textit{enclaves}. Untrusted entities including privileged software (kernel, hypervisor, etc.) cannot access enclave. SGX ensures that the code and data within an enclave cannot be read or modified from outside the enclave.\par 
There are two SGX features that play a vital role in provisioning of sensitive data to an enclave. These are called attestation and sealing.
\begin{itemize}
\item \textit{Attestation: } SGX enclaves are created without privacy-sensitive data. Privacy-sensitive data are delivered after the enclave has been properly instantiated on the platform. The process of demonstrating that a piece of software has been properly instantiated within an enclave on an enabled platform is called \textit{attestation}~\cite{cryptoeprint:2016:1027}. 
\par
Attestation demonstrates to a user that he is communicating with an application running inside an enclave.  This demonstration is accomplished via a cryptographic signature that certifies the hash of the enclave's contents. The remote computer's administrator is able to load any program in an enclave. However, the user (who uses the remote computation service) will deny to load his data into an enclave if the hash of the  contents does not match the desired value~\cite{costanintel}. 
\item \textit{Sealing: } When an enclave is instantiated, SGX provides protections to its data until it is maintained inside the enclave. However, when the enclave process exits, the enclave will be destroyed and all associated data will be lost. If the data is required later, it needs to be stored outside the enclave. Sealing is the process of encrypting and storing data in a way such that only the same enclave would be able un-seal them back to their original form. In our framework, data sealing is not required since the data owners do not necessarily outsource their data to the central server. Instead, they send certain local counts in response to researcher's query. 
\end{itemize}
Memory partition in Intel SGX is described in the supplementary document. 
\subsection{Homomorphic Encryption}
Homomorphic encryption allows performing computation on encrypted data without decrypting the data. The scheme was defined soon after RSA in 1978~\cite{rivest1978data} but was in theory for 30 years. The scheme in a nutshell is: if $c_{1}=\xi(m_{1})$ and $c_{2}=\xi(m_{2})$ (where $m_{1}$ and $m_{2}$ are the plaintexts, $c_{1}$ and $c_{2}$ are the ciphertexts, and $\xi$ is any randomized encryption function), we can perform computation on $c_{1}$ and $c_{2}$ and get the same result as if we were computing with $m_{1}$ and $m_{2}$. 
\par In SAFETY, we have adopted a partial homomorphic system named \textit{Paillier cryptosystem}~\cite{paillier1999public}. Paillier cryptosystem has two important properties that we utilized in SAFETY.
\begin{itemize}
    \item \textit{Probabilistic encryption:} If we encrypt the same message several times using Paillier cryptosystem, it generates different ciphertexts for the same plaintext. 
    \item \textit{Addition homomorphism:} For any public key $n$ and arbitrary messages $m_{1}$, $m_{2}$,
    \begin{align*}
    \xi(m_{1}) + \xi(m_{2}) = (\xi(m_{1}) * \xi(m_{2}))\, mod \, n^{2}
    \end{align*}
    which denotes that we can do an addition operation over ciphertexts.
\end{itemize}

\subsection{Threat Model}\label{threatmodel}
In this paper, our goal is to ensure the confidentiality of individual contributions or data from different geographically distributed data owners. 
Researchers can decrypt only the final result provided by the central server. We also assume that the central server is a semi-honest party (also known as honest-but-curious) where it follows the protocol but may attempt to derive additional information from the server logs or received messages~\cite{goldreich2009foundations}. 
\par
We assume that the computations (required for statistical tests of GWAS) run in an SGX enabled central server. SGX architecture facilitates the central server to perform any computations securely on data provided by multiple data owners. We assume that the processor works properly, and is not compromised. We entirely trust the design and implementation of SGX including all cryptographic operations performed by it. 
\par
It should be mentioned that there is a limited or controlled side-channel attack~\cite{xu2015controlled} proposed against certain SGX based framework~\cite{baumann2015shielding} for a specific scenario. We do not consider such side-channel attacks in this work.


\subsection{Hybrid Approach (SAFETY)}\label{hybrid}
Suppose, there are total $n$ number of data owners ($D_1, D_2, ... , D_n$) connected in the federated environment where a researcher wants to execute a statistical query. The query result should follow or represent as if the query is being executed on the combined dataset. Here, each data owner will have their own individual outputs. For example, data owners $D_1, D_2, \ldots, D_n$ will have outputs $x_1,x_2,\ldots,x_n$ respectively. These outputs can be haplotype or genotype counts (encrypted) for a specific SNP loci based on the query from a researcher. 
\begin{figure}[t]
\centering
\captionsetup{justification=centering,margin=.5cm}
    \includegraphics[width=\textwidth]{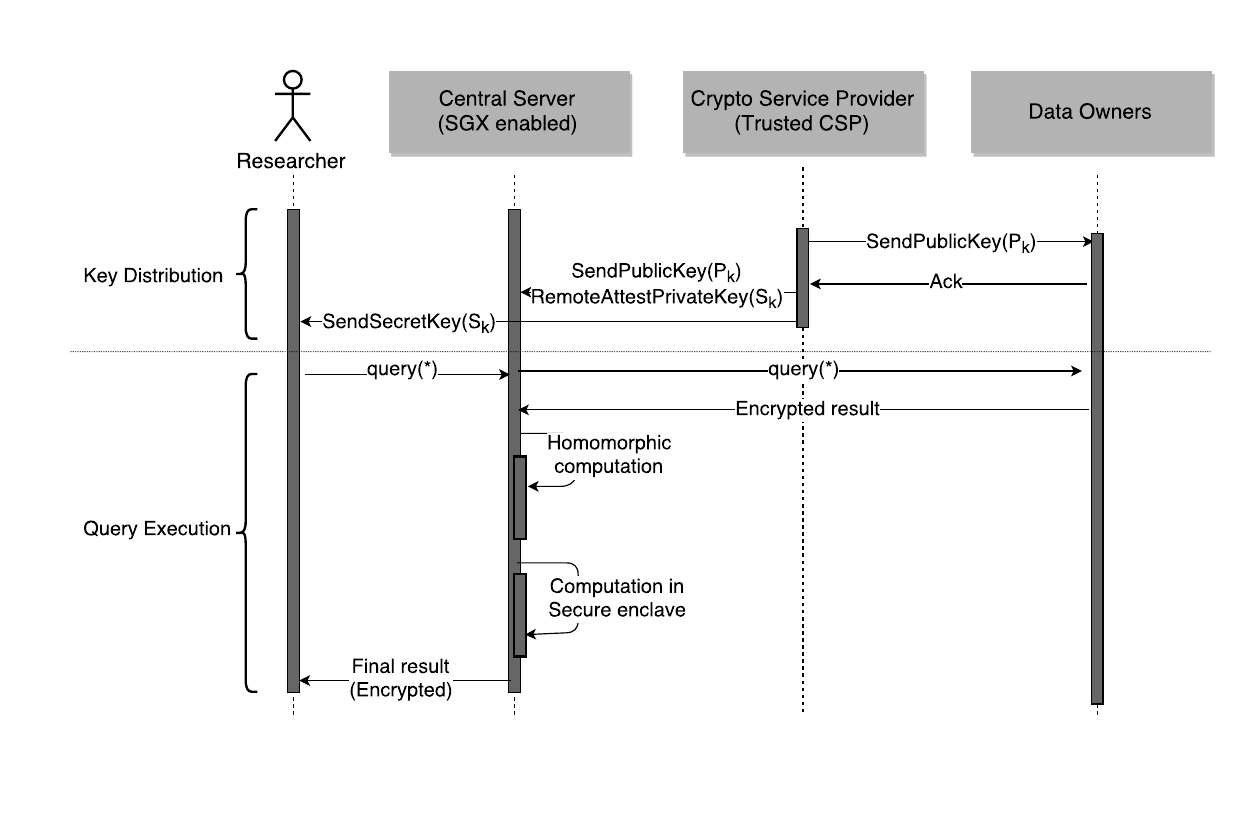}
    \caption{Sequence diagram for the hybrid approach \label{fig:sqdiagram}}
\end{figure}

\par
These outputs are encrypted by the public keys provided beforehand by the CSP. Data owners get the public keys from the CSP before any computation. The data owners generate their encrypted outputs $c_1,c_2,\ldots, c_n$ (from $x_1, x_2, \ldots, x_n$) using the public keys provided by the CSP and send them to the central server for further computations.
\par
The central server then performs homomorphic addition on the individual encrypted outputs $c_1,c_2,\ldots,c_n$ with the Paillier cryptosystem \cite{paillier1999public}. After homomorphic addition, it hands over the total encrypted counts to Intel SGX for further computations required to perform different statistical tests like LD, HWE, CATT, and FET. Then, the total counts are decrypted inside enclave, and further computations are also performed inside enclave where no untrusted application can access these data. The sequence diagram of this protocol is shown in Figure \ref{fig:sqdiagram}.
\par
\begin{figure}[t]
\centering
\captionsetup{justification=centering}
    \includegraphics[width=\textwidth]{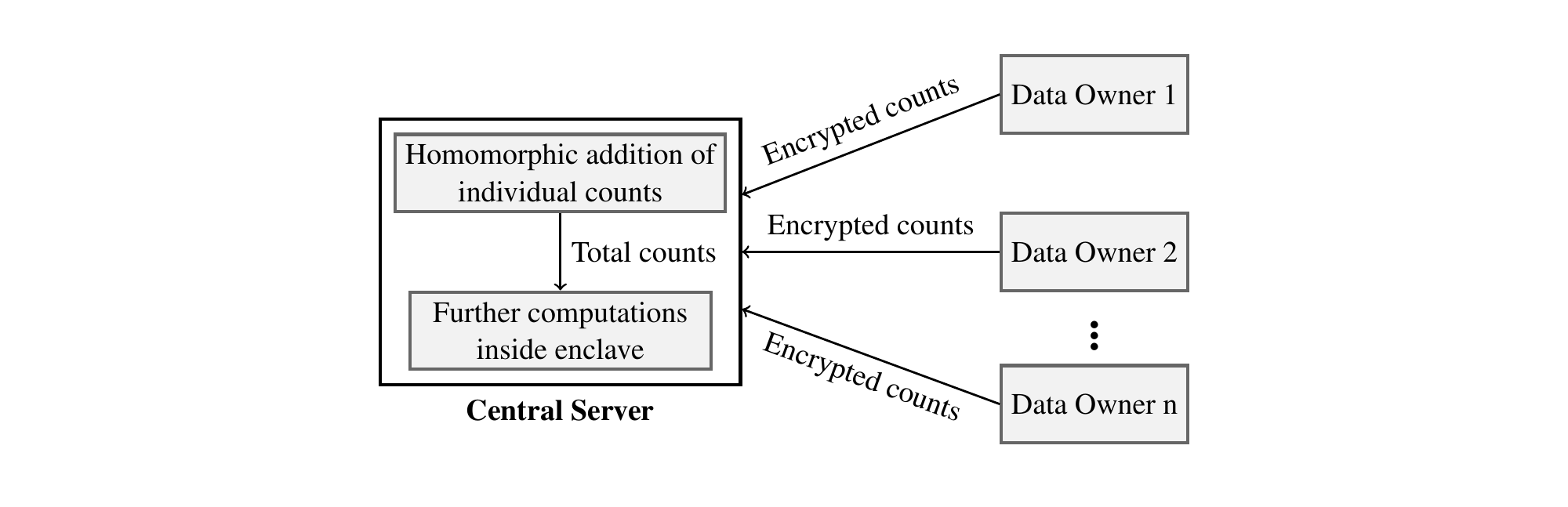}
    \caption{Usage of homomorphic addition in our framework.} \label{fig:homoadditionus}
\end{figure}

It is noteworthy that due to the use of homomorphic addition operation, the number of decryptions required to perform statistical tests is greatly reduced (shown in Table \ref{table:decryptionCount}). Also the individual contributions from the data owners are secured since their values are encrypted and the central server can not learn anything. Figure \ref{fig:homoadditionus} demonstrates the use of homomorphic encryption and Intel SGX in a hybrid architecture. 

\subsection{Secure Hardware Approach}
\label{securehw}
In secure hardware approach, after receiving individual outputs from different data owners, the central server decrypts them inside the enclave and performs further computations on plaintext. The fundamental difference between a hybrid approach and a secure hardware approach is, not using the homomorphic addition on ciphertext. Since in this approach, all the individual encrypted outputs need to be decrypted, the computational overhead is quite large. \\

In the following subsections, we discuss the methods for securely performing LD, HWE, CATT, and FET according to the hybrid approach. 
\\We use the data from Table \ref{table:data_represent} to explain the methods.
\subsection{Secure Linkage Disequilibrium (LD)}
\label{method_ld}
A sample query from researcher regarding LD may look like: \textit{Are rs4305 and rs4630 at linkage disequilibrium}? \\
Both SNPs are bi-allelic. So, there are four possible haplotypes: CA, TA, CG, and TG.
\par 
Each data owners send their haplotype counts which are encrypted by Paillier cryptosystem \cite{paillier1999public}. For instance, data owner 1 sends $N_{CA_{1}}= E(1)$ where the count of \textit{CA} is 1 and $E$ is the encryption function. After receiving the encrypted counts of \textit{CA} from all the data owners, the central server performs homomorphic addition operation on them to obtain the total encrypted count for \textit{CA}. For $n$ data owners,
\begin{align*}
    N_{CA} = N_{CA_{1}}+N_{CA_{2}}+ \ldots +N_{CA_{n}}
\end{align*}
Similarly, total count for TA, CG, and TG are computed. Then the central server instantiates a secure enclave and provisions these encrypted values as an input there. As the decryption key (private key) is sealed by the enclave, it can decrypt the counts and calculate the haplotype frequencies. The haplotype frequencies are calculated in enclave to avoid division of encrypted numbers which is expensive even in fully homomorphic encryption.  Finally, coefficient of the LD is computed and researcher gets the result of his query from this. We discuss detailed procedure for computing LD coefficient in supplementary document.  
\subsection{Secure Hardy-Weinberg Equilibrium (HWE)}
A sample query regarding HWE is: \textit{Does HWE holds at SNP rs4426}? 
\par Possible genotypes at SNP rs4426: CC, CT, and TT. Each data owner will send their individual count for CC, CT, and TT genotypes. After receiving these encrypted genotype counts from all data owners, the central server performs homomorphic addition operation using Paillier cryptosystem \cite{paillier1999public} to obtain total encrypted counts for corresponding genotypes. 
\par Now, all the counts are decrypted inside the enclave to calculate the frequencies $P_{C}$ and  $P_{T}$.
$P_{C}$ is calculated using, 
$P_{C} = \frac{n_{CC}}{n} + \frac{1}{2} \times \frac{n_{CT}}{n}$. 
\par 
Then, $P_{T} = 1 - P_{C}$. So, expected counts for CC, CT and TT are $nP_{C}^{2}, 2nP_{C}P_{T} \text{ and } nP_{T}^{2}$ respectively. Pearson Goodness of Fit Test for HWE is given by:
\begin{equation*}
    \chi^{2} = \frac{(n_{CC} - nP_{C}^{2})^2}{nP_{C}^{2}} + \frac{(n_{CT} - 2nP_{C}P_{T})^2}{2nP_{C}P_{T}} + \frac{(n_{TT} - nP_{T}^{2})^2}{nP_{T}^{2}}
\end{equation*}
Further discussions on HWE are available in supplementary document.
\subsection{Secure Cochran-Armitage Test for Trend (CATT)}
A typical query from researcher regarding CATT is: \textit{Determine if CATT can be inferred at rs4426 ?} \\
Possible genotypes at SNP rs4426 are: CC, CT, and TT. For cases and controls (Cancer positives and negatives respectively), all the data owners send their encrypted genotype counts for both categories to the central server. Homomorphic addition operations are performed to calculate row total and column total using Paillier cryptosystem \cite{paillier1999public}. A contingency table needs to be constructed which is described in the supplementary document. This table is then sent to the enclave where all these row totals and column totals are decrypted for further computations. 

\subsection{Secure Fisher's Exact Test (FET)}
Like CATT, FET also operates on a contingency table. So, for FET, data flow is similar to CATT. Here, the $p-value$ is calculated in enclave after securely aggregating the individual encrypted inputs from the data owners. See supplementary document for further discussions.
\subsection{Pre-computation table for GWAS}
As we have seen, all the statistical tests mentioned before (LD, HWE, CATT, and FET) require processing data in a tabular format. Data owner can keep their data in this format. Consequently, when central server requests for data, data owner can respond readily. It is noteworthy that each data owner has to build the table only once. Thus, pre-computation of the table enhances the efficiency of SAFETY.\\
Table \ref{table:pre_comp3} represents pre-computation table of data owner 1 for performing HWE, CATT, and FET at rs4426.
\begin{table}[ht]
\centering
\caption{Sample pre-computation table for HWE, CATT, and FET}
\label{table:pre_comp3}
\begin{tabular}{|c|c|c|c|c|c|}
\hline
\multicolumn{2}{|c|}{\textbf{CC}} & \multicolumn{2}{c|}{\textbf{CT}} & \multicolumn{2}{c|}{\textbf{TT}} \\ \hline
\textbf{Case}  & \textbf{Control} & \textbf{Case} & \textbf{Control} & \textbf{Case} & \textbf{Control} \\ \hline
1              & 0                & 1             & 0                & 0             & 0                \\ \hline
\multicolumn{2}{|c|}{2}           & \multicolumn{2}{c|}{1}           & \multicolumn{2}{c|}{0}           \\ \hline
\end{tabular}
\end{table}
\par
Since performing LD involves two SNP loci, a different pre-computation table is required. A sample pre-computation table for LD is given in supplementary document.
\section{Discussions}
In this section, we discuss some of the other security and privacy concerns regarding the secure computation of GWAS in our hybrid model.
\subsection{Query Privacy}
In the proposed methods, we do not consider the query privacy of the researcher. In other words, we consider the queries from researcher to be public and data owners, central servers know the targeted position (loci) from the researcher. This issue can be resolved by some of the query privacy or private information retrieval techniques~\cite{schneider2014private,olumofin2010privacy,fung2010privacy}.
\subsection{Output Privacy}
SAFETY does not guarantee the privacy of the final result as that only gets decrypted by researcher. 
We are aware that there are some differential privacy based approaches~\cite{simmons2016enabling, johnson2013privacy, yu2014scalable}, those address this issue and generate differentially private outputs for GWAS. However, as we consider this researcher to be semi-honest, this issue is beyond the scope of the paper.  
\subsection{Consideration of Symmetric Cryptography}
We are not using symmetric cryptography (like AES) for a couple of reasons:
\begin{itemize}
    \item \textbf{Achieving randomized encryption (initialization vector management issue):}  One major drawback of using any symmetric cryptography scheme, (i.e., AES) is achieving randomized or probabilistic encryption. This randomness can be introduced by choosing initialization vectors which needs to be managed by the central server or CSP for multiple data owners. However, SAFETY is based on homomorphic encryption whose encryption is probabilistic by definition, which reduces the burden of managing initialization vectors. 
    
    \item \textbf{Risk of individual contribution leakage:} One of the major concerns in addressing the security of the federated environment is hiding the individual contributions from data owners. As we perform the additions over encrypted data, these contributions are never revealed. As a result, in our proposed framework the possibility of such leakage is highly unlikely. 
    
    \item \textbf{Requires $n$ remote attestations (key distribution problem):} Symmetric cryptography schemes like AES require key distribution/setup with every data owners which results in much network communications which might be prone to attack. On the contrary, our proposed framework is based on public-key cryptography where the data owners use a public key to encrypt their data published by the CSP. As a result, key distribution is much simpler and our framework incurs less communication overhead. 
    
\end{itemize}

\section{Conclusion}
\par Homomorphic encryption and Intel SGX have their own strengths to utilize. Homomorphic encryption can perform some computations without decrypting the ciphertext where Intel SGX can perform any secure computation efficiently after decrypting ciphertext. However, a hybrid model where homomorphic encryption and secure hardware are used in appropriate use cases provides a good trade-off in terms of efficiency and computational support for secure statistical analysis. The outstanding performance of SAFETY attests this hypothesis.
\par Recently, some data analytics and machine learning applications~\cite{schuster2015vc3, ohrimenko2016oblivious,premix} have surfaced adopting Intel SGX for secure computation. However, there are no existing works that use Intel SGX for analyzing genomic data. We think, using secure and efficient computation capability of Intel SGX to analyze genomic data is very promising for healthcare and medical research.

\section*{References}
\bibliographystyle{naturemag}
\bibliography{sample}

\begin{addendum}
\item This work acknowledges funding from NHGRI R00HG008175, NLM R21LM012060, NIBIB U01EB023685, NIGMS R01GM114612, R01GM118574, R01GM118609, NSERC Discovery Grants (RGPIN-2015-04147) and University of Manitoba Startup Grant.

\item[Availability of materials] The evaluation source code can be found at \url{ https://github.com/mominbuet/SafetyGWAS}
\item[Competing Interests.] The authors declare no competing interests.
\item[Contribution.] All authors approved the final manuscript. MNS and MMAA has designed, implemented and evaluated the methods. MNS wrote the majority of the paper and FC, SW, XJ, NM and MMAA provided detailed edits and critical suggestions.
\end{addendum}

\end{document}